# Broken Scale Invariance in the Standard Model[1]

Hitoshi Nishino[2] and Subhash Rajpoot[3]

*Department of Physics & Astronomy*
*California State University*
*1250 Bellflower Boulevard*
*Long Beach, CA 90840*

## Abstract

We introduce Weyl's scale invariance as an additional local symmetry in the standard model of electroweak interactions. An inevitable consequence is the introduction of general relativity coupled to scalar fields *à la* Dirac and an additional vector particle we call the Weylon. We show that once Weyl's scale invariance is broken, the phenomenon (a) generates Newton's gravitational constant $G_{\rm N}$ and (b) triggers spontaneous symmetry breaking in the normal manner resulting in masses for the conventional fermions and bosons. The scale at which Weyl's scale symmetry breaks is of order Planck mass. If right-handed neutrinos are also introduced, their absence at present energy scales is attributed to their mass which is tied to the scale where scale invariance breaks.

---

[1]Work supported in part by NSF Grant # 0308246
[2]E-Mail: hnishino@csulb.edu
[3]E-Mail: rajpoot@csulb.edu

The notion that the standard model [1] is the underlying theory of elementary particle interactions, excluding gravity, is without doubt the prevailing consensus supported by all experiments of the present time. The only missing ingredient is the elusive Higgs particle [2]. It is conceivable that the symmetry breaking mechanism is indeed spontaneous and the Higgs particle will be discovered. However, there are reasons, both aesthetic and otherwise, that necessitate the extensions of the standard model. Grand unification and Proton decay [3] are examples that fall in the former category while neutrino oscillations [4][5] is an example that falls in the latter category.

In this letter we consider extending the standard model with Weyl's local scale invariance [6][7], the doomed symmetry that gave birth to the gauge principle and ultimately paved the way for implementing gauge invariance as we know and practise today. A glance at the elementary particle mass spectrum attests to the fact that scale invariance is a badly broken symmetry of Nature. As we shall show, in the absence of fine-tuning, the scale at which the scale invariance symmetry breaks turns out to be of order Planck mass $M_P \approx 1.3 \times 10^{19}$ GeV. The extended model predicts the existence of an additional vector particle we will call the Weylon. It's mass is tied to the scale at which Weyl's symmetry breaks and is also of order $M_P$.

Implementing scale invariance in the standard model had been previously considered [8][9]. The main result there was the elimination of the Higgs boson from the standard model particle spectrum. The philosophy advocated in the present work is different in spirit. In the present model, the standard model Higgs particle is not eliminated, and is the sought-after particle.

Under scale invariance the parallel transport of a vector around a closed loop in four dimensional space-time not only changes its direction but also its length while the angle between two parallel transported vectors around a closed loop remains the same. The fundamental metric tensor $g_{\mu\nu}$ transforms as

$$g_{\mu\nu}(x) \to \widetilde{g}_{\mu\nu}(x) = e^{2\Lambda(x)} g_{\mu\nu}(x) \;, \tag{1}$$

where $\Lambda(x)$ is the parameter of scale transformations. The four dimensional volume element transforms as

$$d^4x \sqrt{-g} \to e^{4\Lambda(x)} d^4x \sqrt{-g} \;. \tag{2}$$

Since the vierbein $e_\mu{}^m$ and its inverse $e_m{}^\mu$ satisfy $e_\mu{}^m e_{\nu m} = g_{\mu\nu}$ and $e_m{}^\mu e_{n\mu} = \eta_{mn}$ where $(\eta_{mn}) = \text{diag.}\,(1, -1, -1, -1)$ is the tangent space metric, it follows that the transformation properties of $e_\mu{}^m$ and its inverse $e_m{}^\mu$ under Weyl's symmetry are

$$e_\mu{}^m \to e^{\Lambda(x)}\, e_\mu{}^m \;, \quad e_m{}^\mu \to e^{-\Lambda(x)}\, e_m{}^\mu \;. \tag{3}$$

We extend the standard model of particle interactions to include Weyl's scale invariance as a local symmetry. The electroweak symmetry $SU(2) \times U(1)$ is extended to

$$G = SU(2) \times U(1) \times \widetilde{U}(1) \ , \tag{4}$$

where $\widetilde{U}(1)$ represents the local non-compact Abelian symmetry associated with Weyl's scale invariance. The additional particles introduced are the vector boson $S_\mu$ associated with $\widetilde{U}(1)$ and a real scalar field $\sigma$ [10][11][12][13] that transforms as a singlet under $G$. The distinct feature of the new symmetry is that under it fields transform with a real phase whereas under the $SU(2) \times U(1)$ symmetries fields transform with complex phases.

Under $\widetilde{U}(1)$ a generic field in the action is taken to transform as $e^{w\Lambda(x)}$ with a scale dimension $w$. Thus under $G = SU(2) \times U(1) \times \widetilde{U}(1)$ the transformation properties of the entire particle content of the extended model are the following: The $e$-family (g = 1),

$$\Psi_L^{1q} = \begin{pmatrix} u \\ d \end{pmatrix} \sim (2, \tfrac{1}{3}, -\tfrac{3}{2}) \ ; \quad \Psi_L^{1l} = \begin{pmatrix} \nu_e \\ e \end{pmatrix} \sim (2, -1, -\tfrac{3}{2}) \ ;$$

$$\Psi_{1R}^{1q} = u_R \sim (1, \tfrac{4}{3}, -\tfrac{3}{2}) \ ; \quad \Psi_{2R}^{1q} = d_R \sim (1, -\tfrac{2}{3}, -\tfrac{3}{2}) \ ;$$

$$\Psi_{2R}^{1l} = e_R \sim (1, -2, -\tfrac{3}{2}) \ , \tag{5}$$

and similarly for the $\mu$-family (g = 2) and the $\tau$-family (g = 3). All of these fermions have the same scale dimension $w = -3/2$. The scalar bosons comprising the Higgs doublet $\Phi$ and the real scalar $\sigma$,

$$\Phi \sim (2, -1, -1) \ ; \quad \sigma \sim (1, 0, -1) \ , \tag{6}$$

with the common scale dimension $w = -1$. We introduce $W_\mu$, $B_\mu$ and $S_\mu$ as the gauge potentials respectively associated with the $SU(2), U(1), \widetilde{U}(1)$ symmetries. We suppress the $SU(3)$ of strong interactions as neglecting it will not affect our results and conclusions. The action $I$ of the model is

$$\begin{aligned} I = \int d^4x \sqrt{-g} \bigg[ & -\tfrac{1}{4} g^{\mu\rho} g^{\nu\sigma} (W_{\mu\nu} W_{\rho\sigma} + B_{\mu\nu} B_{\rho\sigma} + U_{\mu\nu} U_{\rho\sigma}) \\ & + \sum_{\substack{f=q,l \\ g=1,2,3 \\ i=1,2}} \left( \overline{\Psi}_L^{\mathrm{gf}} e_m{}^\mu \gamma^m D_\mu \Psi_L^{\mathrm{gf}} + \overline{\Psi}_{iR}^{\mathrm{gf}} e_m{}^\mu \gamma^m D_\mu \Psi_{iR}^{\mathrm{gf}} \right) + g^{\mu\nu} (D_\mu \Phi)(D_\nu \Phi^\dagger) + \tfrac{1}{2} g^{\mu\nu} (D_\mu \sigma)(D_\nu \sigma) \\ & + \sum_{\substack{f=q,l \\ g,g'=1,2,3 \\ i=1,2}} \left( \mathbf{Y}_{gg'}^{f} \overline{\Psi}_L^{\mathrm{gf}} \Phi \Psi_{iR}^{\mathrm{gf}} + \mathbf{Y}'^{f}_{gg'} \overline{\Psi}_L^{\mathrm{gf}} \widetilde{\Phi} \Psi_{iR}^{\mathrm{g'f}} \right) + \text{h.c.} - \tfrac{1}{2} (\beta \phi^\dagger \Phi + \zeta \sigma^2) \widetilde{R} + V(\Phi, \sigma) \bigg] \ , \end{aligned} \tag{7}$$

where $\widetilde{\Phi} \equiv i\sigma_2 \phi^*$, the indices (g, g') are for generations, the indices f = (q, l) refer to (quark, lepton) fields, $\mathbf{Y}_{gg'}^{f}$ or $\mathbf{Y}'^{f}_{gg'}$ are quark, lepton Yukawa couplings that define the mass matrices after symmetry breaking, the index $i = 1, 2$ is needed for right-handed fermions, while $\beta$ and $\zeta$ are dimensionless couplings. The various $D$'s acting on the fields represent the covariant derivatives constructed in the usual manner using the principle of minimal substitution. Explicitly,

$$\begin{aligned}
D_\mu \Psi_L^{\text{gf}} &= \left(\partial_\mu + ig\tau \cdot W_\mu + \tfrac{i}{2}g' Y_L^{\text{gf}} B_\mu - \tfrac{3}{2} f S_\mu - \tfrac{1}{2}\tilde{\omega}_\mu{}^{mn}\sigma_{mn}\right)\Psi_L^{\text{gf}} \;, \\
D_\mu \Psi_{iR}^{\text{gf}} &= \left(\partial_\mu + \tfrac{i}{2}g' Y_{iR}^{\text{gf}} B_\mu - \tfrac{3}{2} f S_\mu - \tfrac{1}{2}\tilde{\omega}_\mu{}^{mn}\sigma_{mn}\right)\Psi_{iR}^{\text{gf}} \;, \\
D_\mu \Phi &= \left(\partial_\mu + ig\tau \cdot W_\mu - \tfrac{1}{2}g' B_\mu - f S_\mu\right)\Phi \;, \\
D_\mu \sigma &= \left(\partial_\mu - f S_\mu\right)\sigma \;.
\end{aligned} \qquad (8)$$

The $Y_L^{\text{gf}}$'s, $Y_{iR}^{\text{gf}}$'s represent the hypercharge quantum numbers (e.g., f = q, g = 1, i = 1, $Y_L^{1q} = 1/3$, $Y_{1R}^{1q} = 4/3$, etc.), $g$, $g'$, $f$ are the respective gauge couplings of $SU(2)$, $U(1)$, $\tilde{U}(1)$, while

$$U_{\mu\nu} \equiv \partial_\mu S_\nu - \partial_\nu S_\mu \qquad (9)$$

is the field strength associated with Weyl's $\tilde{U}(1)$. It is gauge invariant, since $S_\mu$ transforms as

$$S_\mu \to S_\mu - \tfrac{1}{f}\partial_\mu \Lambda \;. \qquad (10)$$

The spin connection $\tilde{\omega}_\mu{}^{mn}$ [14] is defined in terms of the vierbein $e_\mu{}^m$

$$\begin{aligned}
\tilde{\omega}_{mrs} &\equiv \tfrac{1}{2}(\tilde{C}_{mrs} - \tilde{C}_{msr} + \tilde{C}_{srm}) \;, \\
\tilde{C}_{\mu\nu}{}^r &\equiv (\partial_\mu e_\nu{}^r + f S_\mu e_\nu{}^r) - (\partial_\nu e_\mu{}^r + f S_\nu e_\mu{}^r) \;,
\end{aligned} \qquad (11)$$

while the affine connection $\tilde{\Gamma}^\alpha{}_{\mu\nu}$ is defined by

$$\tilde{\Gamma}^\rho{}_{\mu\nu} = \tfrac{1}{2}g^{\rho\sigma}\left[(\partial_\mu + 2 f S_\mu)g_{\nu\sigma} + (\partial_\nu + 2 f S_\nu)g_{\mu\sigma} - (\partial_\sigma + 2 f S_\sigma)g_{\mu\nu}\right] \;. \qquad (12)$$

The Riemann curvature tensor $\tilde{R}^\rho{}_{\sigma\mu\nu}$ is

$$\tilde{R}^\rho{}_{\sigma\mu\nu} = \partial_\mu \tilde{\Gamma}^\rho{}_{\nu\sigma} - \partial_\nu \tilde{\Gamma}^\rho{}_{\mu\sigma} - \tilde{\Gamma}^\lambda{}_{\mu\sigma}\tilde{\Gamma}^\rho{}_{\nu\lambda} + \tilde{\Gamma}^\lambda{}_{\nu\sigma}\tilde{\Gamma}^\rho{}_{\mu\lambda} \;, \qquad (13)$$

where $\tilde{\Gamma}^\rho{}_{\mu\nu}$, $\tilde{R}^\rho{}_{\sigma\mu\nu}$ and the Ricci tensor $\tilde{R}^\rho{}_{\mu\rho\nu} = \tilde{R}_{\mu\nu}$ have scale dimension $w = 0$, while the scalar curvature $\tilde{R} = g^{\mu\nu}\tilde{R}_{\mu\nu}$ has the form

$$\begin{aligned}
\tilde{R} &= R - 6 f D_\mu S^\mu + 6 f^2 S_\mu S^\mu \;, \\
D_\kappa S^\mu &= \partial_\kappa S^\mu + \tilde{\Gamma}^\mu{}_{\kappa\nu} S^\nu \;,
\end{aligned} \qquad (14)$$

and transforms with scale dimension $w = -2$. The potential $V(\phi, \sigma)$ is given by

$$V(\Phi, \sigma) = \lambda \, (\Phi^\dagger \Phi)^2 - \mu \, (\Phi^\dagger \Phi) \, \sigma^2 + \xi \, \sigma^4 \quad , \tag{15}$$

where $\lambda$, $\mu$, $\xi$ are dimensionless couplings. It is interesting to note that the scalar potential in this model consists of quartic terms only as required by Weyl's scale invariance. Yet the desired descent, a two stage process, of $G$ to $U(1)_{\text{em}}$

$$G = SU(2) \times U(1) \times \widetilde{U}(1) \rightarrow SU(2) \times U(1) \rightarrow U(1)_{\text{em}} \tag{16}$$

is possible. In the primary stage of symmetry breaking, scale invariance symmetry is broken. This is achieved by setting

$$\sigma(x) = \tfrac{1}{\sqrt{2}} \Delta \quad , \tag{17}$$

where $\Delta$ is a constant for the symmetry breaking scale associated with Weyl's $\widetilde{U}(1)$. The primary stage of symmetry breaking also determines Newton's gravitational constant $G_{\text{N}}$,

$$\zeta \Delta^2 = \frac{1}{4\pi G_{\text{N}}} \quad . \tag{18}$$

Thus $\Delta \approx 0.3 \times M_{\text{P}}/\sqrt{\zeta}$ and barring any fine-tuning $\Delta \approx \mathcal{O}(M_{\text{P}})$, if we take $\zeta \approx \mathcal{O}(1)$. At this stage the scalar field $\sigma$ becomes the goldstone boson [15][16]. The vector particle associated with $\widetilde{U}(1)$ breaking, the Weylon, absorbs the goldstone field and becomes massive with mass $M_{\text{S}}$ given by

$$M_{\text{S}} = \sqrt{\frac{3 f^2}{4\pi G_{\text{N}}}} \approx 0.5 \times f M_{\text{P}} \quad . \tag{19}$$

Thus $M_{\text{S}} \approx \mathcal{O}(M_{\text{P}})$ in the absence of fine-tuning $f \approx \mathcal{O}(1)$. Weyl's $\widetilde{U}(1)$ symmetry decouples completely and the scalar potential after the primary stage of symmetry breaking takes the form

$$V(\Phi) = -\mu \Delta^2 (\Phi^\dagger \Phi) + \lambda \, (\Phi^\dagger \Phi)^2 + \frac{\xi}{4} \Delta^4 \quad . \tag{20}$$

It is to be noted that this form of the potential, apart from the vacuum energy density term contributing to the cosmological constant, is of the same form as the standard Higgs potential in the standard model. All the conventional particles are still massless at this stage. With $G_{\text{N}}$ defined, it is appropriate to work in the weak field approximation. Henceforth we set $\sqrt{g} g_{\mu\nu} \approx \eta_{\mu\nu} + \mathcal{O}(\kappa)$ where $\kappa^2 = 16\pi G_{\text{N}}$. The secondary stage of symmetry breaking is spontaneous. This takes place when $\Phi \rightarrow \langle \Phi \rangle$ where

$$\langle \Phi \rangle = \tfrac{1}{\sqrt{2}} \begin{pmatrix} \eta \\ 0 \end{pmatrix} \quad , \tag{21}$$

$$\eta = \sqrt{\frac{\mu \Delta^2}{\lambda}} \ , \tag{22}$$

and $\eta$ is the electroweak symmetry breaking scale of order 250 GeV. In the standard model, $\mu$ and $\lambda$ are unrelated while in this model they are related,

$$\frac{\mu}{\lambda} = \left(\frac{\eta}{\Delta}\right)^2 \approx 2.4 \times \zeta \, G_F^{-1} M_P^{-2} \approx 10^{-33} \times \zeta \ .$$

After spontaneous symmetry breaking (SSB), the conventional particles acquire masses as in the standard model,

$$M_W = \tfrac{1}{2} g \eta \ , \qquad M_Z = \frac{M_W}{\cos \theta_W} \ ,$$
$$\mathbf{M}^f_{gg'} = \tfrac{1}{\sqrt{2}} \mathbf{Y}^f_{gg'} \eta \ , \qquad \mathbf{M'}^f_{gg'} = \tfrac{1}{\sqrt{2}} \mathbf{Y'}^f_{gg'} \eta \ , \tag{24}$$

where $\theta_W$ is the weak angle and $\mathbf{M}^f_{gg'}, \mathbf{M'}^f_{gg'}$ are the quark ($f = q$) and the charged lepton ($f = l$) mass matrices. At this stage neutrinos are still massless. In this model there is still left over the conventional Higgs particle $h_0$ with mass given by

$$M_{h_0} = \sqrt{\mu} \Delta \approx 0.3 \times \sqrt{\frac{\mu}{\zeta}} \, M_P \ , \tag{25}$$

which is undetermined as $\mu$ and $\zeta$ are still free parameters. It is interesting to note that in this model the mass of the Higgs particle is tied to the scale associated with the breaking of Weyl's $\widetilde{U}(1)$ symmetry which is of order Planck mass. In principle, $M_{h_0}$ can be as large as $M_P$ posing problems with unitarity. However, although the standard model is a renormalizable theory [17][18], the present model is not. This puts into doubt the validity of the unitarity constraint derived in the renormalizable standard model and extrapolated to the non-renormalizable extended model considered here. After SSB, the mass of the Weylon gets shifted,

$$M_S \to \sqrt{\frac{3f^2}{4\pi G_N}\left(1 + \frac{\beta \eta^2}{\zeta \Delta^2}\right)} \ . \tag{26}$$

However, the additional contribution is negligibly small as $\eta^2/\Delta^2 \approx 10^{-33}$. Apart from being superheavy, another distinct property of the Weylon is that it completely decouples from the fermions and the bosons of the standard model.

At the present time, one fundamental issue is that of neutrino masses and their lightness as compared to the masses of other particles. In the standard model and the model under consideration, neutrinos are strictly massless as no right-handed neutral lepton fields were introduced. A popular extension of the standard model that addresses this issue in an aesthetically appealing way introduces right-handed neutrinos $\Psi^{1l}_{1R} = \nu_{eR}$, $\Psi^{2l}_{1R} = \nu_{\mu R}$, $\Psi_{1R}^{3l} = \nu_{\tau R}$ that lead to seesaw masses [19] for the the conventional neutrinos. This scenario is usually entertained in the $SO(10)$ grand unified theory, where the right-handed neutrinos acquire super heavy masses. The super heavy scale is determined by the stage at which the internal symmetry $SO(10)$ breaks, and has nothing to do with gravitational interactions. If right-handed neutrino fields are also introduced in the present model, the seesaw mechanism can naturally be accommodated due to the presence of the singlet field $\sigma$. The relevant interaction Lagrangian is

$$L_\nu = \sum_{\substack{g,g'=1,2,3 \\ i=1}} \left( \mathbf{Y}_{gg'}^l \overline{\Psi}_L^{gl} \Phi \Psi_{iR}^{g'l} + \text{h.c.} + \frac{1}{2} \mathbf{Y}_{gg'}^{RR} \sigma_{1R}^{gl\,T} C \sigma \Psi_{1R}^{g'l} \right) \quad . \tag{27}$$

Lepton number is explicitly broken by the last term. Scale breaking gives superheavy Majorana masses to the right-handed neutrinos and SSB subsequently gives Dirac masses that connects the left- and right-handed neutrinos leading to the following familiar $6 \times 6$ mass matrix

$$\mathbf{M}_\nu = \frac{1}{\sqrt{2}} \begin{pmatrix} \mathbf{0} & \mathbf{Y}_{gg'}^l \eta \\ \mathbf{Y}_{g'g}^l \eta & \mathbf{Y}_{gg'}^{RR} \Delta \end{pmatrix} , \tag{28}$$

the eigenvalues of which are three seesaw masses for the light neutrinos and three heavy neutrinos with enough parameters to fit the observed solar and atmospheric neutrino oscillation phenomena. In the present model, the scale of right-handed neutrino masses is tied to the scale $\Delta$ associated with Weyl's $\widetilde{U}(1)$ breaking which in turn is tied to Newton's constant $G_N$. This is unlike the GUT scenario where right-handed neutrino masses are tied to the GUT scale at which the grand unification internal symmetry breaks. Thus the absence of right-handed neutrinos from the low energy scales is attributed to their superheavy masses of $\mathcal{O}(M_P)$, and may be interpreted as indication that right-handed neutrinos (and also gauge-mediated right-handed currents) and gravitational interactions may ultimately be related.

We stress that our model needs only quartic potential (15) for the scalar fields $\Phi$ and $\sigma$ only with dimensionless couplings as its foundation. The scale-breaking parameter $\Delta$ then induces the quadratic terms in the resulting potential (20). Whereas in the standard model $\mu$ and $\lambda$ are not related, our model relates them in terms of $\Delta$ *via* (22).

We note that the symmetry breaking scheme depicted in the model under consideration would apply universally to theories that accommodate local scale invariance and generate Newton's constant $G_N$ as a symmetry breaking effect. SSB necessarily requires the scalar potential to contain terms quadratic in scalar fields. Such terms are either added explicitly or generated via quantum corrections [20]. In scale invariant theories the scalar potential consists of terms only quartic in the scalar fields. Thus in GUT theories with both local scale invariance and internal symmetry invariance, it is a scale invariance breaking that would precede spontaneous symmetry breaking. This is because since all such theories would contain the scalar curvature $R$, Newton's constant $G_N$ would be generated as the primary symmetry breaking effect. After scale breaking, the resulting potential would contain the necessary terms quadratic in scalar fields to effect SSB, similar to the discussion in the text, resulting in the GUT scale $M_G$, intermediate scale(s) $M_I$ ($M_I$, $M_{II}$, $M_{III}$, $\cdots$) and the electroweak scale $M_W \approx \sqrt{G_F^{-1}}$ with the hierarchy $M_G > M_I > M_{II} > M_{III} > \cdots > M_W$.

Our contention is that the present model presents a viable scheme in which gravity is unified, albeit in a semi-satisfactory way, with the other interactions. In the standard model physical fields and the couplings like electric charge $e = 1/\sqrt{g^{-2} + g'^{-2}}$ and Fermi constant $G_F = g^2/(8M_W^2)$ get defined *after* SSB. Similarly, in the present model, not only $e$ and $G_F$, but also $G_N$ gets defined *after* symmetry breaking, thus conforming to the main theme in physics that all phenomena observed in Nature are symmetry breaking effects. When the complete theory of all interactions is found, the model in its present form, it is hoped, will serve as its low energy limit.

To conclude, we have accommodated Weyl's scale invariance as a local symmetry in the standard electroweak model. This inevitably leads to the introduction of general relativity. The additional particles are one vector particle we call the Weylon and a real scalar singlet that couples to the scalar curvature $\widetilde{R}$ *à la* Dirac. The scale at which Weyl's scale invariance breaks defines Newton's gravitational constant $G_N$. Weyl's vector particle, *i.e.*, the Weylon absorbs the scalar singlet $\sigma$ and acquires mass $\mathcal{O}(M_P)$ in the absence of fine tuning. The scalar potential is unique in the sense that it consists of terms only quartic in the scalar fields and dimensionless couplings. Yet, as we have demonstrated, symmetry breaking is possible such that the left-over symmetry is $U(1)_{em}$ and all particle masses are consistent with present day phenomenology. If right-handed neutrinos are also introduced, the light neutrinos acquire seesaw masses and the suppression factor in the neutrino masses is of $\mathcal{O}(M_P)$.